# A Partitioning Methodology for Accelerating Applications in Hybrid Reconfigurable Platforms *

M.D. Galanis[1], A. Milidonis[1], G. Theodoridis[2], D. Soudris[3], and C.E. Goutis[1]

[1] VLSI Design Lab., Electrical & Computer Engineering Department, University of Patras, Rio, Greece
[2] Section of Electronics & Computers, Physics Department, Aristotle University, Thessalonica, Greece
[3] VLSI Design Center, Electrical & Computer Eng. Dept., Democritus University, Xanthi, Greece
e-mail: mgalanis@ee.upatras.gr

**Abstract**

*In this paper, we propose a methodology for partitioning and mapping computational intensive applications in reconfigurable hardware blocks of different granularity. A generic hybrid reconfigurable architecture is considered so as the methodology can be applicable to a large number of heterogeneous reconfigurable platforms. The methodology mainly consists of two stages, the analysis and the mapping of the application onto fine and coarse-grain hardware resources. A prototype framework consisting of analysis, partitioning and mapping tools has been also developed. For the coarse-grain reconfigurable hardware, we use our previous-developed high-performance coarse-grain data-path. In this work, the methodology is validated using two real-world applications, an OFDM transmitter and a JPEG encoder. In the case of the OFDM transmitter, a maximum clock cycles decrease of 82% relative to the ones in an all fine-grain mapping solution is achieved. The corresponding performance improvement for the JPEG is 43%.*

## 1. Introduction

Reconfigurable architectures have been a topic of intensive research activities in the past years. Reconfigurable fabrics are able to unify the performance of ASICs and the flexibility offered by the microprocessors [1]. In particular, hybrid (mixed) granularity reconfigurable systems [1]-[4] offer extra advantages in terms of performance and great flexibility to implement efficiently computational intensive applications (like DSP and multimedia) characterized by mixed functionality (data and control). Such hybrid architectures usually consist of: fine-grain reconfigurable units usually implemented in FPGA technology, coarse-grain reconfigurable units implemented in ASIC technology, data and program memories, reconfigurable interconnection network, and microprocessor(s). Due to the special characteristics of the heterogeneous (coarse and fine-grain) reconfigurable units included in a hybrid platform, certain parts of the application are better suited to be executed on the coarse-grain units and other parts on the fine-grain units.

The fine-grain reconfigurable hardware's granularity usually ranges from 1 to 2-bits and it is realized by an embedded FPGA unit. Small bit-width operations can be efficiently executed by fine-grain hardware, as the granularity of the CLBs of the embedded FPGA is typically one or two bits. Tasks of Finite State Machine (FSM) type of functionality (i.e. control structures) are also good candidates to be implemented by the fine-grain reconfigurable hardware. The coarse-grain reconfigurable blocks are implemented in ASIC technology and execute the word-level or sub word-level operations. These blocks can slightly modify their functionality according to the application requirements.

It is widely adopted that the execution of word-level operations by coarse-grain units offers great advantages in terms of delay, area and reconfiguration time compared to the execution of these operations by the fine-grain reconfigurable units [1]. So, to exploit these advantages, the development of a methodology for partitioning an application in two parts, where the one is executed in the coarse-grain reconfigurable hardware and the other one in the fine-grain one, and mapping efficiently these parts on the corresponding reconfigurable units, is required.

In this paper, a formalized and automated partitioning methodology is presented for hybrid reconfigurable systems. The methodology is parameterized with respect to the reconfigurable hardware, i.e. the fine and the coarse-grain parts of the target architecture. It is assumed that both types of reconfigurable hardware are characterized in terms of timing and area characteristics. The introduced methodology was developed for the purposes of a European IST project, called Architectures and Methodologies for Dynamic REconfigurable Logic (AMDREL) [5]. The project's consortium is composed by industrial and academic partners.

The main contributions of the methodology are the analysis procedure at the basic block level of the application and the mapping procedures to the fine and coarse-grain hardware. A prototype software framework has been developed for implementing the proposed methodology. The methodology is evaluated in this paper using two real-world applications, an IEEE 802.11a OFDM transmitter and a JPEG encoder, both developed by the industrial partners of the AMDREL project [5]. The work presented here is an extension of our previous work [6], where we developed a high-performance coarse-grain data-path and a methodology for mapping applications onto it.

---

* This work is partially supported by the project IST-2001-34793-AMDREL funded by the E.C.
Also, it was partially funded by the Alexander S. Onassis Public Benefit foundation.

The considered generic reconfigurable platform, which mainly targets the DSP and multimedia domains, is shown in Figure 1. The platform includes coarse and fine-grain reconfigurable hardware units for data processing, shared data memory, and a reconfigurable interconnection network. All the above components are typically integrated in a System-on-Chip (SoC) configured by microprocessor(s). This generic architecture can model a variety of existing hybrid (mixed) reconfigurable architectures, like the ones presented in [2], [3], [4].

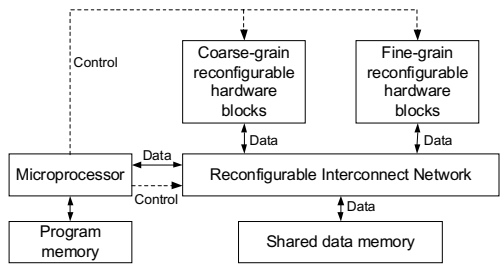

**Figure 1. Generic reconfigurable platform architecture.**

The paper is organized as follows: The related work is presented in section 2, while section 3 describes the proposed partitioning methodology. Experimental results are given in section 4. Finally, section 5 concludes this paper and presents future activities.

## 2. Related work

There has been considerable research for developing reconfigurable architectures in the past [1]. We are focusing on hybrid granularity platforms containing fine and coarse-grain reconfigurable hardware blocks.

The Pleiades [3] architecture is an approach that combines an on-chip microprocessor with a number of heterogeneous reconfigurable units of different granularities connected via a reconfigurable interconnection network. The Strategically Programmable System (SPS) [2] is a hybrid reconfigurable system architecture that combines fine-grain reconfigurable units and ASIC coarse-grain modules which are pre-placed within a fully reconfigurable fabric. Chameleon project [4] considers a hybrid reconfigurable platform that contains microprocessor(s), an FPGA unit and coarse-grain reconfigurable units, called Field Programmable Function Arrays (FPFAs).

In the aforementioned hybrid reconfigurable systems, there is no a formalized partitioning methodology between the fine and coarse-grain reconfigurable units. In [4], a rather empirical manner is followed, which assigns bit-level parts of an application to the FPGA.

There has been also work in hardware-software partitioning of applications for reconfigurable architectures consisting of one RISC-type microprocessor and FPGA(s) [7], [8]. However, those works do not consider coarse-grain reconfigurable blocks, thus they can not benefit from the computational ability of these units [2], [3], [4].

## 3. Partitioning methodology

Research activities [9] have shown that basic blocks inside loop structures represent a significant portion of the execution time for DSP and multimedia applications. The term *basic block* expresses a sequence of instructions (operations) with no branches into or out of the middle. At the end of each basic block there is a branch that controls which basic block executes next. The proposed partitioning methodology focuses on finding the most critical basic blocks (called *kernels*) of the input application. These kernels are executed on the coarse-grain hardware, so that the total execution time of the application meets the specification's requirements. With the term critical, we define the basic blocks that represent computationally intensive parts of the application. The critical basic blocks are often located in nested loops.

We are interested in exploiting the available parallelism at the operation level (either in critical or in non-critical parts of an application) in both types of reconfigurable hardware. Therefore, the methodology supports mutually exclusive execution of the fine and the coarse-grain hardware. According to their original C/C++ specification, critical and non-critical parts of the application are executed purely sequentially in the coarse and the fine-grain part of the architecture, respectively. Although, the methodology supports the mutual exclusive operation, this does not imply that either the fine or the coarse-grain hardware remain unused during the execution of the application. This is due to the fact that our generic hybrid reconfigurable architecture targets DSP and multimedia applications. Typically these applications process certain amount of data (called frames) whose computation is repeated over time. Through the pipelining among the stages of computations, the reconfigurable processing units of the hybrid architecture are always utilized.

In the following, we explain the proposed partitioning methodology. In Figure 2, the flow of the proposed methodology is shown. The input is the application (or a part of the application) which is described in a high-level language like C/C++. The application's source code is considered as the output of a hardware/software partitioning stage, which defines the parts that they are going to be executed in the reconfigurable hardware.

In the *first step*, the Control Data Flow Graph (CDFG) representation is created from the source code. This model of computation is extensively used in mapping applications on reconfigurable hardware. The CDFG is the input in the steps of mapping to fine and coarse-grain hardware and in the partitioning engine. In *step 2*, the application is mapped to the fine-grain hardware and the execution time is calculated. If the overall execution time of the application meets the timing constraints, then the methodology exits, since there is no need to continue with the next steps, i.e. to partition the application into fine and coarse-grain hardware. If the timing constraints are not satisfied, then we proceed to *step 3*, which is the analysis procedure.

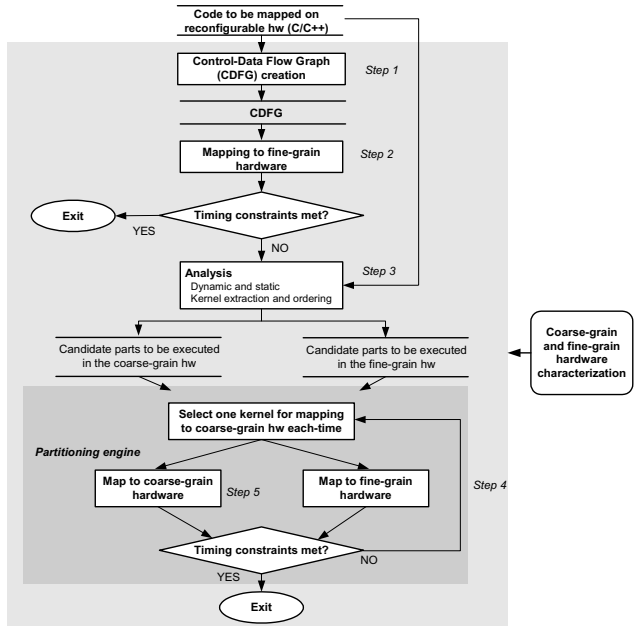

**Figure 2. Block diagram of the proposed partitioning methodology.**

In the analysis procedure (*step 3*), the application's source code is processed, so as to identify the dominant kernels, which are the candidates to be mapped to the coarse-grain hardware. The rest of the application code is mapped to the fine-grain part of the architecture. The identification of kernels is a combination of *dynamic* and *static* analysis. The kernels are ordered in decreasing order of computational complexity.

In the partitioning engine (*step 4*), kernels are moved one by one for execution (and thus acceleration) in the coarse-grain hardware. After the movement of each kernel in the coarse-grain part, the total execution time of the application is calculated to check if the timing constraints are met. To compute the execution time, the mapping to the fine and coarse-grain hardware takes place. The time required for communicating data values through the shared data memory of Figure 1, between the two types of hardware is also taken into account.

The mapping to the coarse-grain part of the architecture is the *step 5* of the proposed methodology. If there is still a violation in the overall execution time of the application, the procedures of moving kernels to the coarse-grain hardware and mapping the parts of the application onto the fine and coarse-grain parts, are repeated until the timing constraints are satisfied. As the mapping procedures to both types of reconfigurable hardware determine the execution time, we have developed appropriate algorithms to perform these procedures.

### 3.1 Analysis step

The analysis step is the procedure for identifying the kernels of the input application and provides the input to the partitioning engine block, as it is shown in Figure 2. More specifically, it identifies the critical and non-critical parts of the application. The critical part is the set of kernels, which are the basic blocks inside loops that cause performance overheads. These kernels are candidates to be mapped to the coarse-grain hardware, while the non-critical parts of the application are executed in the fine-grain hardware.

The inherent computational complexity (counts of basic operations and memory accesses) is a meaningful measure to identify dominant kernels. This information can be obtained through a combination of: (a) dynamic analysis (profiling), and (b) static analysis within basic blocks of the input specification. Since operations in a basic block do not have a uniform cost, a weighted sum is calculated and aggregated at the basic block level to indicate the computational complexity within the application. The weights indicate the delay allocated to each basic operator.

For the *dynamic* analysis, the source code is executed with appropriate input and profiling information is gathered at the basic block level. For performing dynamic analysis, we have used Lex [10], which a lexical analyzer used for parsing the input code. By developing the proper scripts in Lex, we can identify loop (*for*, *while* and *do-while*) and conditionals (*if-then-else*) structures in the source code. Then, Lex automatically places a counter for each basic block in a loop. The modified source code (after counter placement) is compiled and executed with the input vectors that represent the typical operation of the application. The placed counter gives the access count for each basic block of the input program.

Lex is also used for the *static* analysis. It identifies the basic operations and the memory accesses inside the basic blocks and generates a detailed and illustrative overview of the distribution of the algorithm complexity over basic operators. The total weight (complexity) of a basic block is computed as the product of the basic block execution frequency (*exec_freq*) times the weight of the operations of this basic block (*bb_weight*), i.e.:

$$total\_weight = exec\_freq \cdot bb\_weight \quad (1)$$

After all critical basic blocks have been identified, an ordering of these critical basic blocks takes place. These kernels are sorted in descending order of computational complexity. Thus, the first kernel which is going to be mapped onto the coarse-grain hardware, if the overall execution requirement is not met, is the most computational intensive one.

### 3.2 Mapping to fine-grain hardware

The proposed mapping methodology for the fine-grain part of the architecture is a temporal partitioning algorithm. The temporal partitioning resolves the hardware implementation of an application that does not fit into the fine-grain reconfigurable hardware by time-sharing the hardware in a way that each partition fits in the available resources (for example the CLBs of an FPGA). This time-sharing of the hardware is achieved through the dynamic reconfiguration of the device, which is the case in contemporary FPGA devices, either commercial [11] or academic ones.

The mapping methodology classifies the nodes in the Data Flow Graph (DFG) of the input application according to their As Soon As Possible (ASAP) levels [12]. The ASAP levels expose the parallelism hidden in the DFG, i.e. all the DFG nodes with the same level can be considered for parallel execution without any dependency check. There also exists some degree of parallelism among the nodes with different levels, i.e. if they are not connected by a data edge. The approach followed is that the nodes are executed in increasing order relative to their ASAP levels. This ensures stable inputs for every DFG node at the next ASAP level. The mapping methodology also handles CDFG, by iteratively mapping the DFGs composing the CDFG.

The pseudocode of the proposed mapping algorithm to the fine-grain hardware is illustrated in Figure 3. *Partition*($u_i$) denotes the temporal partition to which the node $u_i$ belongs ($1 \leq u_i \leq N$, $N$ is the number of DFG nodes) and *max_level* denotes the maximum ASAP level of any node in the DFG. The algorithm traverses each node of the DFG, level by level, and assigns them to a partition. The DFG nodes are assigned to partitions numbered 1 and beyond. All the nodes from level 1 to *max_level* are traversed. Nodes of the same ASAP level are placed in a single partition and if the available area in the fine-grain hardware is exhausted then the nodes are assigned to the next partition. If the nodes in the current ASAP level are all assigned to a partition, then the next level nodes are considered. Initially, a partition has no nodes.

The $A_{FPGA}$ is the area available for mapping the DFG operations in the fine-grain (FPGA) reconfigurable hardware. To ensure that the routing of the resources is feasible, the $A_{FPGA}$ is a percentage of the total FPGA area. A typical value is a 70% of the overall FPGA area. The $size(u_i)$, which is the area occupied by the mapped DFG node, and the $A_{FPGA}$ are dependent from the fine-grain technology (e.g. a specific FPGA device [11]). Since these are parameters in our methodology, the proposed methodology is applicable to every type of reconfigurable fine-grain hardware.

```
i = 1;
level = 1;
area_covered = 0;
while(level ≤ max_level)
  for each node u_i with level(u_i)= level
    current_area = size(u_i);
    if (area_covered + current_area ≤ A_FPGA)
      partition(u_i) = i;
      area_covered = area_covered + current_area;
    end if;
    else
      i = i + 1;
      partition(u_i) = i;
      area_covered = current_area;
    end else;
    level = level + 1;
  end for;
end while;
```

**Figure 3. Mapping algorithm to fine-grain hardware.**

The shared data memory of the hybrid reconfigurable platform (Figure 1) is used for storing the input and output values among the temporal partitions. For each temporal segment a configuration bit-stream is generated. According to the application's data- and control-flow, the appropriate configuration bit-stream is loaded to the FPGA device. For each temporal partition, full reconfiguration of the fine-grain hardware is performed. Thus, the reconfiguration time has the same value for each partition and it is added to the execution time of each temporal partition.

### 3.3 Mapping to coarse-grain hardware

For the coarse-grain hardware, the high-performance coarse-grain data-path and the mapping methodology presented in [6] are considered. This data-path consists of a set of Coarse-Grain Components (CGCs) implemented in ASIC technology, a reconfigurable interconnection network, and a register bank. The CGC is an *n×m* array of nodes, where *n* is the number of rows and *m* the number of columns. The connections among the CGC nodes are reconfigured by appropriate steering logic. This allows to easily realize any complex operations (like a multiply-add operation) and increase system's performance [6]. Each CGC node contains a multiplier and ALU where only one of them is activated in a clock cycle.

Due to the flexible structure of the CGC-based data-path, any required computational structure can be easily implemented; thus the CGC data-path can realize the behaviour of any existing coarse-grain data-path, like the ones in [2], [3]. This is the main reason why we have considered in this work this specific type of coarse-grain hardware. Also, due to the CGC data-path's features, the stages of the mapping methodology are accommodated by simple, yet efficient algorithms. The steps of the mapping process are: (a) scheduling of DFG operations, and (b) binding with the CGCs. A proper list-based scheduler has been developed. After CGC binding [6], the overall latency of the DFG is measured in clock cycles having period $T_{CGC}$.

This period is set for having unit execution delay for the CGCs. For handling CDFG, the mapping procedure is iterated through the DFGs comprising the CDFG of an application.

### 3.4 Partitioning engine

The partitioning engine moves kernels one by one to the coarse-grain hardware until the performance requirements are satisfied. After the movement of each kernel to the coarse-grain hardware, the total execution time of the application is calculated to check if the timing constraints are met. The mapping procedures to the fine and coarse-grain hardware are required for computing the execution time. If the timing constraints are not met, the process of moving kernels to the coarse-grain hardware is repeated until the timing constraints are satisfied. The total execution time is:

$$t_{total} = t_{FPGA} + t_{coarse} + t_{comm} \quad (2)$$

where $t_{FPGA}$ is the execution time in the FPGA (fine-grain) hardware, $t_{coarse}$ is the execution time in the coarse-grain data-path and $t_{comm}$ is the time required for transferring data between the two types of reconfigurable hardware, through the shared data memory (Figure 1).

The $t_{coarse}$ equals to:

$$t_{coarse} = \sum_i t_{to\_coarse}(BB_i) \cdot Iter(BB_i) \quad (3)$$

where $t_{to\_coarse}$ is the time required for executing the basic block $BB_i$ in the coarse-grain hardware, and $Iter(BB_i)$ is the number of times that the $BB_i$ is invoked. Similarly, $t_{FPGA}$ equals to:

$$t_{FPGA} = \sum_i t_{to\_FPGA}(BB_i) \cdot Iter(BB_i) \quad (4)$$

The $t_{FPGA}$ includes the reconfiguration time for all the generated temporal partitions after the mapping of the basic blocks.

### 4. Experimental results

We have developed a prototype framework in C++ to implement the flow of Figure 2. For the software development, we have also used academic open source tools. For example, the SUIF2 [13] and MachineSUIF [14] compiler infrastructures has been used and proper passes have been developed for the CDFG creation. As already mentioned, Lex has been used for the dynamic and static analysis.

In this paper, we apply the proposed partitioning methodology to two applications written in C language by the AMDREL's partners [5]. The first one is the front-end of the baseband processing of an IEEE 802.11a OFDM transmitter. The front-end consists of the Quadrature Amplitude Modulation (QAM) unit, the IFFT block and the cyclic prefix unit. The considered source code of the OFDM transmitter is composed by 18 basic blocks (BBs). The second application is a JPEG encoder. The main parts of the JPEG encoder are the DCT transformation unit, the quantizer, the zig-zag scanning unit and the entropy (Huffman) encoder. The considered JPEG encoder source code consists of 22 BBs.

Table 1 reports the total weights of both applications (last column), in decreasing order of value, of the 8 most computational intensive basic blocks, extracted by the analysis step of the partitioning methodology. The second column of this Table reports the execution frequency of the specific basic block, while the third column gives the weights of the operations in the basic block. The execution frequency values are taken for a number of 6 payload symbols for the OFDM transmitter, while for the JPEG encoder for transforming an image of size $256 \cdot 256$ bytes. These inputs of the applications also hold for the clock cycles results of Table 2 and 3. The DFGs of the basic blocks of both applications consist of arithmetic operations of type ALU and multiplication; thus no divisions are present in the DFGs. In the analysis process, we give a weight equal to 1 for the ALU operations and a weight equal to 2 for the multiplication ones, since the later ones have a larger computational complexity.

**Table 1. Ordered total weights of basic blocks**

| Basic Block no. | Basic Block exec. freq. | Operations weight | Total weight |
|---|---|---|---|
| *OFDM transmitter* | | | |
| 22 | 336 | 115 | 38640 |
| 12 | 1200 | 25 | 30000 |
| 3 | 864 | 6 | 5184 |
| 5 | 370 | 12 | 4440 |
| 42 | 800 | 5 | 4000 |
| 32 | 560 | 6 | 3360 |
| 29 | 448 | 7 | 3136 |
| 21 | 147 | 18 | 2646 |
| *JPEG encoder* | | | |
| 6 | 355024 | 3 | 1065072 |
| 2 | 8192 | 85 | 696320 |
| 1 | 8192 | 83 | 679936 |
| 22 | 65536 | 5 | 327680 |
| 8 | 30927 | 8 | 247416 |
| 3 | 65536 | 3 | 196608 |
| 16 | 63540 | 3 | 190620 |
| 17 | 63540 | 2 | 127080 |

Table 2 and 3 show the results of the partitioning methodology, in terms of clock cycles, after using the developed partitioning framework for the OFDM transmitter and the JPEG encoder. The following assumptions hold for both applications. The clock cycle period is set to the clock period of the fine-grain (FPGA) hardware. We have considered that the clock cycle period of the FPGA hardware is three times larger than the CGC data-path's clock period, i.e. $T_{FPGA} = 3 \cdot T_{CGC}$. This is a rather moderate assumption for the performance gain of an ASIC technology compared to an FPGA one. For these experiments, two values of $A_{FPGA}$ are considered: 1500 and 5000 units of area. For each case of the $A_{FPGA}$, the coarse-grain data-path consists of two and three 2x2 CGCs. Thus, four different cases are considered in this experiment. For all of these cases, a timing constraint of 60000 clock cycles has to be satisfied for the OFDM transmitter, while for the JPEG encoder a timing constraint of $11 \cdot 10^6$ clock cycles has to be met.

Table 2. OFDM partitioning results for timing constraint of 60000 clock cycles

|  | $A_{FPGA}$=1500 | | $A_{FPGA}$=5000 | |
|---|---|---|---|---|
| Initial Cycles | 263408 | | 124080 | |
| CGCs no. | two 2x2 | three 2x2 | two 2x2 | three 2x2 |
| Cycles in CGC | 53184 | 41472 | 53184 | 41472 |
| BB no. | 22, 12, 3 | 22, 12 | 22, 12, 3 | 22, 12 |
| Final cycles | 57088 | 47856 | 56864 | 46512 |
| % cycles reduction | 78.3 | 81.8 | 54.1 | 62.5 |

Table 3. JPEG partitioning results for timing constraint of $11 \cdot 10^6$ clock cycles

|  | $A_{FPGA}$=1500 | | $A_{FPGA}$=5000 | |
|---|---|---|---|---|
| Initial Cycles ($\cdot 10^6$) | 18434 | | 12399 | |
| CGCs no. | two 2x2 | three 2x2 | two 2x2 | three 2x2 |
| Cycles in CGC ($\cdot 10^6$) | 5817 | 5699 | 5817 | 5669 |
| BB no. | 6, 2, 1 | 6, 2, 1 | 6, 2, 1 | 6, 2, 1 |
| Final cycles ($\cdot 10^6$) | 10558 | 10411 | 10423 | 10227 |
| % cycles reduction | 42.7 | 43.5 | 15.9 | 17.5 |

The first row of Table 2 and 3 shows the number of cycles for an all-FPGA implementation of the considered applications. It is evident that an all-FPGA solution cannot satisfy the timing constraint. The third row of Table 2 and 3 shows the clock cycles required for the implementation of the BBs (their number is shown in the fourth row of the Table 2 and 3) when these are mapped to the CGC data-path. The number and type of CGCs are given in the second row. The BBs of the fourth row have been chosen from the partitioning methodology for execution on the coarse-grain.

The final clock cycles, after the partitioning, are shown in the fifth row of Table 2 and 3. It is clear from these results, that by choosing costly BBs to be mapped in the coarse-grain reconfigurable hardware, system's performance is largely improved and the timing constraint is satisfied. These results prove the effectiveness of both the proposed partitioning methodology and the automated framework. Also, as the FPGA area grows, the reduction of clock cycles is smaller since a larger FPGA exploits better the parallelism of an application due to the considered fine-grain mapping algorithm shown in Figure 3. A maximum clock cycles reduction of approximately 82% relative to the all-FPGA solution of the OFDM transmitter, is reported for the case of $A_{FPGA}$=1500 and three 2x2 CGCs present in the CGC data-path.

## 5. Conclusions - Future work

A methodology for partitioning applications between fine and coarse-grain reconfigurable blocks of a hybrid granularity architecture, was presented. We also gave specific mapping algorithms for the fine and coarse-grain reconfigurable blocks. The experiments showed that the timing constraints of an application can be satisfied by proper functional partitioning. On going-work considers multiple threads of execution for parallel operation of the fine and coarse-grain reconfigurable blocks. Future work focuses on partitioning an application for satisfying energy consumption constraints.